\documentclass[showpacs,preprint,showkeys,superscriptaddress,amssymb,
               nofootinbib,aps]{revtex4}

\makeatletter
\renewcommand{\p@subsection}{}
\renewcommand{\p@subsubsection}{}
\usepackage{array,multirow}
\usepackage{mdwlist}
\usepackage{amsmath}
\usepackage{amsthm}

\usepackage{amssymb}
\usepackage{bm}
\usepackage{blindtext}
\usepackage{enumitem}
\usepackage{graphicx}
\def\ni{\noindent}
\def\be{\begin{equation}}
\def\ee{\end{equation}}
\def\ba{\begin{eqnarray}}
\def\ea{\end{eqnarray}}

\begin{document}
\title{\Large Perfect fluid warp drive solutions\\with the cosmological
      constant}
\author{Osvaldo L. Santos-Pereira}\email{olsp@if.ufrj.br}
\affiliation{Physics Institute, Universidade Federal do Rio de
            Janeiro, Rio de Janeiro, Brazil}
\author{Everton M.\ C.\ Abreu}\email{evertonabreu@ufrrj.br}
\affiliation{Physics Department, Universidade Federal Rural do
	Rio de Janeiro, Serop\'edica, Brazil}
\affiliation{Physics Department, Universidade Federal de Juiz de
             Fora, Juiz de Fora, Brazil}
\affiliation{Multidisciplinary Graduate Program in Applied Physics, 
	Physics Institute, Universidade Federal do Rio de Janeiro,
	Rio de Janeiro, Brazil}
\author{Marcelo B. Ribeiro}\email{mbr@if.ufrj.br}
\affiliation{Physics Institute, Universidade Federal do Rio de
            Janeiro, Rio de Janeiro, Brazil}
\affiliation{Multidisciplinary Graduate Program in Applied Physics, 
	Physics Institute, Universidade Federal do Rio de Janeiro,
	Rio de Janeiro, Brazil}
\date{\today}
\begin{abstract}
\ni The Alcubierre metric describes a spacetime geometry that allows
a massive particle inside a spacetime distortion, called warp bubble,
to travel with superluminal global velocities. In this work we advance
solutions of the Einstein equations with the cosmological constant for
the Alcubierre warp drive metric having the perfect fluid as source.
We also consider the particular dust case with the cosmological constant,
which generalizes our previous dust solution (Santos-Pereira et al.\
2020) and led to vacuum solutions connecting the warp
drive with shock waves via the Burgers equation, as well as our perfect
fluid solution without the cosmological constant (Santos-Pereira et al.\
2021). All energy conditions are also analyzed. The
results show that the shift vector in the direction of the warp bubble
motion creates a coupling in the Einstein equations that requires
off-diagonal terms in the energy-momentum source. Therefore, it seems
that to achieve superluminal speeds by means of the Alcubierre warp drive
spacetime geometry one may require a complex configuration and
distribution of energy, matter and momentum as source in order to produce
a warp drive bubble. In addition, warp speeds seem to require more
complex forms of matter than dust for stable solutions and that negative
matter may not be a strict requirement to achieve global superluminal
speeds.
\end{abstract}
\pacs{04.20.Gz; 04.90.+e; 47.40.-x}
\keywords{warp drive, cosmic fluid, Burgers equation, shock waves,
          Alcubierre geometry}
\maketitle
\section{Introduction}
\renewcommand{\theequation}{1.\arabic{equation}}
\setcounter{equation}{0}

The \textit{warp drive} is a mechanism based on General Relativity which
in theory allows for massive particles to be propelled throughout the
spacetime with global superluminal speeds \cite{Alcubierre1994,
Alcubierre2017}. The theory describes this possibility by means
of a localized spacetime distortion, called \textit{warp bubble}, that
would contain a local lightcone where the particle would follow special
relativity, that is, move locally with speeds smaller than light. However,
the metric is such that the warp bubble moves along a geodesic that creates
an expansion of spacetime behind it and a contraction in front of it in
such a way that an observer outside the warp bubble sees it moving with
superluminal speed. 

In the original paper by M.\ Alcubierre the warp drive metric was
established without solving the Einstein equations \cite{Alcubierre1994}.
The Einstein tensor components were calculated and it was noticed that
negative energy density would be required to create the warp bubble,
thus violating the weak and dominant energy conditions.

Ford and Roman \cite{FordRoman1996} calculated via quantum inequalities
the amount of negative energy required for the warp drive to be possible,
concluding that it would be a massive amount, impossible to achieve.
Pfenning and Ford \cite{Pfenning1997} also concluded that it would be 
necessary an enormous amount of energy for the warp drive to be possible.
They obtained a quantity ten orders of magnitude greater than the
mass-energy of the entire  visible Universe, also with negative density.

Krasnikov \cite{Krasnikov1998} discussed the possibility of a massive
particle moving in space faster than a photon, arguing that this is 
not possible due to limitations on globally hyperbolic spacetimes 
properties with feasible physical assumptions. He is the creator of
a specific spacetime topology with devices that would allow massive
particles to travel between two points in space with superluminal
velocities without the need for tachyons. Everet and Roman
\cite{EveretRoman1997} coined the name for this spacetime topology as
the \textit{Krasnikov tube}. They generalized the metric designed by
Krasnikov by proposing a tube in the direction of the particle's path,
connecting the start and end point. Inside this tube, the spacetime is
flat and the lightcones are opened to allow the one direction
superluminal travel. The Krasnikov tube also requires huge amounts of
negative energy density. Since the tube is designed to not possess
closed timelike curves, it would be theoretically possible to construct
a two way non-overlapping system that could work as a time machine. The
energy-momentum tensor (EMT) for the Krasnikov metric is positive in
some regions. Both the metric and the obtained EMT were thoroughly
analyzed in Refs.\ \cite{Lobo2002,Lobo2003}.

Van de Broeck \cite{Broeck1999} made a relevant contribution to warp 
drive theory by demonstrating that a small modification of the original
Alcubierre geometry would reduce, to a few solar masses, the total
negative energy necessary for the creation of the warp bubble distortion
of spacetime. This result have led van de Broeck to suppose that other
geometrical modifications of the Alcubierre's geometry for the warp
drive could also reduce the amount of energy necessary to create a warp
drive bubble in the same way. 

Natario \cite{Natario2002} stated that the spacetime contraction and
expansion of the warp bubble is a peculiar consequence of the warp
drive metric.  Hence, he designed a spacetime where no contraction or
expansion occurs for the warp drive bubble. Lobo and Visser
\cite{LoboVisser2004b, LoboVisser2004} discussed that the center of the
warp bubble proposed by Alcubierre needs to be massless \cite[see also
Refs.][]{White2003, White2011}. They proposed a linearized model for both
Alcubierre and Natario proposals and demonstrated that for small speeds,
the energy stored in the warp fields must be a significant fraction of
the mass of the spaceship inside the warp bubble. Quarra \cite{quarra}
discussed null geodesics moving faster-than-light according to far away
observers when inside a region-delimited gravitational wave field. Lee
and Cleaver \cite{cleaver1,cleaver2} analyzed how external radiation
might affect the Alcubierre warp bubble to turn it unsustainable. They
also claimed that a warp field interferometer could not detect spacetime
distortions. Mattingly et al.\ \cite{cleaver3,cleaver4} studied the
curvature invariants characteristic of Natario and Alcubierre warp drives,
whereas Mattingly \cite{matt} discussed further curvature invariants for
warped spacetimes.

Bobrick and Martire \cite{bobrick} claimed that any warp drive spacetime
consists of a shell of regular or exotic material moving inertially with
a certain speed, also reaching at a class of subluminal spherically
symmetric warp drives. Lentz \cite{lentz} and Fell and Heisenberg
\cite{fell} advanced superluminal capable soliton solutions with positive
energy warp drives. Santiago et al.\ \cite{santiago,santiago2} argued that only
comoving timelike Eulerian observers satisfy the weak energy condition,
whereas this not the case for all timelike observers. Furthermore, they
claimed that all physically reasonable warp drives will violate the null
and weak energy conditions, therefore disputing the claim advanced by
Refs.\ \cite{bobrick,lentz,fell} that it would be theoretically possible
to set up positive energy warp drives that satisfy the weak energy
condition. 

Motivated by the fact that neither the original paper by Alcubierre,
nor the subsequent ones cited above, did actually solve the Einstein
equations using the warp drive metric, we proceed to investigate
possible solutions for a dust particle energy momentum tensor \cite{nos}.
Our results showed that solutions of the Einstein equations for the
Alcubierre warp drive metric having dust as source connect them in a
particular case the warp drive geometry to the well-known Burgers
equation, which describes the dynamics of the waves moving through an
inviscid fluid. Hence, shock waves appear to be vacuum solutions of the
Einstein equations endowing the warp drive metric \cite{nos}.

In our second paper \cite{nos2} we investigated solutions for the warp
drive metric having the perfect fluid and a special case of anisotropic
fluid with heat flux, but both with zero cosmological constant in the
Einstein equations. The resulting solutions indicate that positive
matter densities are possibly capable of generating superluminal speeds. 
In our third paper \cite{nos3} a charged dust was used as source EMT for
the Alcubierre metric and the Einstein equations which included the
cosmological constant. We obtained solutions connecting the electric
energy density with the cosmological constant and, again, some solutions
were found having positive matter density and satisfying the energy
conditions. 

Motivated by the results we obtained in Ref.\ \cite{nos3} we have
pondered that even though the Alcubierre warp drive metric is a vacuum
geometry, the warp bubble would be created by geometry alone, or if a
vacuum energy would make it possible through other material sources of
energy and momentum. Hence, in this paper we went back to the perfect
fluid source but included the cosmological constant in the Einstein
equations as an additional flexibility and geometrical properties for 
the solutions.
 
We calculated the Einstein equations and analyzed the null divergence 
of the energy momentum tensor together with the validity requirements 
for the energy conditions inequalities to be satisfied. We found that
the perfect fluid with the cosmological constant as source for the
Alcubierre warp drive results in four sets of differential equations,
two of them are very similar and raise the possibility for the shift
vector to be a complex function in one case, depending on the $(t,y)$
coordinates, and in another case depending on the $(t,z)$ coordinates.
The other two sets of solutions are identical to each other and similar
to the solution we found in Ref.\ \cite{nos}, except that now there is
a cosmological constant coupled with the Burgers equation and, again,
the warp drive is connected to shock waves solutions. Considering that
the zero pressure reduces the perfect fluid to the dust EMT, the
solution for this case is identical to the one we found in Ref.\
\cite{nos}, namely, the vacuum solution of the Einstein equations
connecting the warp drive to shock waves.

The plan of the paper is as follows. Section 2 presents a brief review
of the basic equations and concepts of the warp drive theory, and in
Section 3 we solve the Einstein equations and calculated the covariant
divergence for the EMT. Section 4 discusses the energy conditions
inequalities and their validity for the warp drive with $\Lambda\not=0$
and the perfect fluid as a source. In section 5 we analyze incoherent
matter as a source assuming that this is as a special case considering
the perfect fluid with null pressure. In section 6 we depict our
conclusions and final remarks.

\section{Warp drive basic concepts}
\renewcommand{\theequation}{2.\arabic{equation}}
\setcounter{equation}{0}

\subsection{Warp drive metric}
 
The warp drive metric is a generic metric in a foliated spacetime 
given by the following expression,
\be
{ds}^2 = - \left(\alpha^2 -\beta_i\beta^{i}\right) \, dt^2 
+ 2 \beta_i \, dx^i \, dt + \gamma_{ij} \, dx^i \, dx^j \,\,,
\label{wdmetric1}
\ee
where $d\tau$ is the  proper time lapse, $\alpha$ is the lapse function
that controls the amount of time elapsed between two hypersurfaces of
constant time coordinate, $\beta^i$ is the spacelike shift vector and
$\gamma_{ij}$ is the spatial metric for the hypersurfaces. The lapse
function $\alpha$ and the shift vector $\beta_i$ are functions of the
spacetime coordinates to be determined, $\gamma_{ij}$ is a positive-definite 
metric on each one of the spacelike hypersurfaces and these features make
this spacetime globally hyperbolic. Throughout this paper Greek indices will 
range from 0 to 3, whereas the Latin ones indicate the spacelike 
hypersurfaces and will range from 1 to 3.

We have the following choices for Eq.\ \eqref{wdmetric1}
\cite{Alcubierre1994},
\begin{align}
\alpha &= 1, 
\\
\beta_1& = - v_s(t)f\big[r_s(t)\big], \label{betax}
\\
\beta_2 &= \beta_3 = 0,
\\
\gamma_{ij} &= \delta_{ij}.
\end{align}
Hence, the \textit{warp drive metric} is given by,
\be
ds^2 = - \left[1 - v_s(t)^2 f(r_s)^2\right]dt^2 
- v_s(t) f(r_s)\,dx\,dt 
+ dx^2 + dy^2 + dz^2 \,\,,
\label{alcmetric1}
\ee
where $v_s(t)$ is the velocity of the center of the bubble moving
along the curve $x_s(t)$, given by,
\be
v_s(t) = \frac{dx_s (t)}{dt}\,\,.
\ee 
The function $f(r_s)$ is the warp drive \textit{regulating form 
function}. It describes the shape of the warp bubble, which is 
given by the expression,
\be
f(r_s) = \frac{\tanh\left[\sigma(r_s + R)\right] 
- \tanh\left[\sigma(r_s - R)\right]}
{2 \tanh(\sigma R)} \,\,,
\label{regfunction}
\ee
where $\sigma$ and $R$ are constants to be determined. The function 
$r_s(t)$ defines the distance from the center of the bubble
$[x_s(t),0,0]$ to a generic point $(x,y,z)$ on the surface of the 
bubble, given by the following equation,
\be
r_s(t) = \sqrt{\left[x - x_s(t)\right]^2 + y^2 + z^2}\,\,.
\label{bubbleradius}
\ee 
From Eq. \eqref{bubbleradius} one can see that the warp bubble 
is perturbed in a one-dimensional manner because of the term 
$x - x_s(t)$.

\subsection{Einstein tensor components}

The components of the Einstein tensor with a cosmological constant 
for the warp drive metric in Eq. \eqref{wdmetric1} are given by 
the expressions below,
\be 
G_{00} =  \Lambda(1-\beta^2) - \frac{1}{4} 
(1 + 3\beta^2)
\left[
\left(\frac{\partial \beta}{\partial y} \right)^2 +  
\left(\frac{\partial \beta}{\partial z} \right)^2 
\right] 
- \beta \left(\frac{\partial^2 \beta}{\partial y^2} + 
\frac{\partial^2 \beta}{\partial z^2}\right),
\label{et00}
\ee

\be
G_{01} = \Lambda \beta + \frac{3}{4} 
\beta \left[
\left(\frac{\partial \beta}{\partial y}\right)^2 
+ \left(\frac{\partial \beta}{\partial z}\right)^2 
\right] 
+ \frac{1}{2}\left(
\frac{\partial^2 \beta}{\partial y^2} 
+ \frac{\partial^2 \beta}{\partial z^2}
\right),
\label{et01}
\ee

\be
G_{02} = - \frac{1}{2}
\frac{\partial^2 \beta}{\partial x \partial y} 
- \frac{\beta}{2} 
\left(2\frac{\partial \beta}{\partial y}
\, \frac{\partial \beta}{\partial x} +
\beta \frac{\partial^2 \beta}{\partial x \partial y} +
\frac{\partial^2 \beta}{\partial t \partial y}\right),
\label{et02}
\ee

\be
G_{03} = - \frac{1}{2}
\frac{\partial^2 \beta}{\partial x \partial z} 
- \frac{\beta}{2} 
\left(2\frac{\partial \beta}{\partial z}
\, \frac{\partial \beta}{\partial x} +
\beta \frac{\partial^2 \beta}{\partial x \partial z} +
\frac{\partial^2 \beta}{\partial t \partial z}\right),
\label{et03}
\ee

\be
G_{11} = \Lambda - \frac{3}{4} \left[
\left(\frac{\partial \beta}{\partial y}\right)^2 
+ \left(\frac{\partial \beta}{\partial z}\right)^2
\right], \label{et11}
\ee

\be
G_{12} = \frac{1}{2}\left(
2 \frac{\partial \beta}{\partial y} \, 
\frac{\partial \beta}{\partial x} 
+ \beta \frac{\partial^2 \beta}{\partial x \partial y} 
+ \frac{\partial^2 \beta}{\partial t \partial y}\right),
\label{et12}
\ee

\be
G_{13} = \frac{1}{2}\left(
2 \frac{\partial \beta}{\partial z} \, 
\frac{\partial \beta}{\partial x} 
+ \beta \frac{\partial^2 \beta}{\partial x \partial z} 
+ \frac{\partial^2 \beta}{\partial t \partial z}\right),
\label{et13}
\ee

\be
G_{23} = \frac{1}{2} \frac{\partial \beta}{\partial z} 
\, \frac{\partial \beta}{\partial y},
\label{et23}
\ee

\be
G_{22} = - \Lambda - \frac{1}{4}\left[
\frac{\partial^2 \beta}{\partial t \partial x}
+ \beta \frac{\partial^2 \beta}{\partial x^2}
+ \left(\frac{\partial \beta}{\partial x}\right)^2
\right]
- \frac{1}{4}\left[
\left(\frac{\partial \beta}{\partial y}\right)^2
- \left(\frac{\partial \beta}{\partial z}\right)^2
\right],
\label{et22}
\ee

\be
G_{33} = - \Lambda - \frac{1}{4}\left[
\frac{\partial^2 \beta}{\partial t \partial x}
+ \beta \frac{\partial^2 \beta}{\partial x^2}
+ \left(\frac{\partial \beta}{\partial x}\right)^2
\right]
+ \frac{1}{4}\left[
\left(\frac{\partial \beta}{\partial y}\right)^2
- \left(\frac{\partial \beta}{\partial z}\right)^2
\right]\,\,, 
\label{et33}
\ee
where $\beta = - \beta_1 = 
v_s(t)f(r_s)$, as in Eq.\,\eqref{betax}. Also noticed that we 
incorporated the cosmological constant into the Einstein tensor
\be
G_{\mu\nu} \to G_{\mu\nu} - \Lambda g_{\mu\nu}
\ee

\subsection{Energy conditions revisited}

The components for the Eulerian (normal) observers' 4-velocities are
given by,
\be
u^\alpha = \left(1, - \beta , 0, 0\right), \ \ 
u_\alpha = (- 1,0,0,0)\,\,.
\label{eulerianvel}
\end{equation}
Using these results into the Einstein equations,
\be
T_{\alpha \beta} u^\alpha u^\beta = \frac{1}{8\pi}G_{\alpha \beta} 
u^\alpha u^\beta\,\,,
\label{energdens}
\ee 
results in an expression concerning the energy conditions.
From Eqs.\,\eqref{eulerianvel} and considering that the only non-zero terms
of Eq.\ (\ref{energdens}) are $G_{00}$, $G_{01}$ and $G_{11}$, we 
obtain the following expression,
\be
T_{\alpha \beta} \, u^\alpha u^\beta 
= \frac{1}{8\pi}\left(G_{00} - 2 \beta G_{01} 
+ \beta^2 G_{11}\right).
\label{endenalc}
\ee
Substituting Eqs.\,\eqref{et00}, \,\eqref{et01} and \eqref{et11} into
Eq.\ \eqref{endenalc} the result may be written as,
\be
T_{\alpha \beta} \, u^\alpha u^\beta 
= \Lambda - \frac{v_s^2}{32 \pi}
\left[
\left(\frac{\partial f}{\partial y} 
\right)^2 +  \left(\frac{\partial f}{\partial z} 
\right)^2
\right].
\label{edwd1}
\ee
The bubble radius is given by using Eq. \eqref{bubbleradius}. So,
Eq.\ (\ref{edwd1}) is given by,
\be
T_{\alpha \beta} u^\alpha u^\beta  
= \Lambda - \frac{v_s^2}{16 \pi}\frac{y^2 + z^2}{r_s^2}
\left(\frac{\partial f}{\partial r_s}\right)^2.
\label{edwd2}
\ee
This result is similar to the one found by Alcubierre \cite{Alcubierre1994},
with the difference that Ref.\ \cite{Alcubierre1994} did not consider the
cosmological constant. Considering the results in Ref.\
\cite{Alcubierre1994} we realized that both the weak and dominant energy
conditions would be violated \cite{nos} if the bubble was formed. However,
these same energy conditions would be satisfied in the case of a vacuum
solution, which discloses the new result that the warp drive metric is a
vacuum solution for the Einstein equations. Besides, the Burgers equation
is connected to this geometry where shock waves are partial solutions. Here,
with the inclusion of the cosmological constant it may be possible that
the weak and dominant energy conditions could be satisfied if $\Lambda$ is
positive and large enough in Eq. \eqref{edwd2}.

\section{Matter content energy-momentum tensors}
\renewcommand{\theequation}{3.\arabic{equation}}
\setcounter{equation}{0}

\subsection{Perfect fluid energy momentum tensor}

For Eulerian observers 4-velocity $u^\alpha = \left(1,-\beta,0,0\right)$
and $u_\alpha = (- 1,0,0,0)$ the perfect fluid EMT for those observers is
given by the expression below,
\be
T_{\alpha \beta} = (\mu + p) \, u_{\alpha} u_{\beta} + p g_{\mu \nu}\,\,,
\ee
where $\mu$ is a scalar function that represents the matter density, 
$p$ is the fluid pressure, and $g_{\mu\nu}$ is the metric tensor. 
It must be noted that the dust EMT is a particular case for the perfect
fluid with null pressure.

From the Einstein tensor components, Eqs.\ \eqref{et00} to \eqref{et33},
and the perfect fluid EMT it is possible to write all the components of
Einstein equations. After some algebraic work we found the following set
of equations, 
\be
\frac{4}{3} \Lambda = 8 \pi \left[T_{00} + 2 \beta T_{01} 
+ \left(\beta^2 - \frac{1}{3} \right)T_{11}\right] = 
8 \pi \left(\mu - \frac{1}{3}p\right)\,,	
\label{eqset1}
\ee

\be
\left(\frac{\partial\beta}{\partial y}\right)^2 +
\left(\frac{\partial\beta}{\partial z}\right)^2 
- 4 \Lambda = - 32 \pi\left(T_{00} + 2 \beta T_{01} 
+ \beta^2 T_{11}\right) = - 32 \pi \mu \,\,,
\label{eqset2}
\ee

\be
\frac{\partial^2 \beta}{\partial y^2} + 
\frac{\partial^2 \beta}{\partial z^2} = 
16 \pi (T_{01} + \beta T_{11}) = 0\,,
\label{eqset3}
\ee

\be
\left(\frac{\partial \beta}{\partial y}\right)^2 
+ \left(\frac{\partial \beta}{\partial z}\right)^2 
- \frac{4}{3} \Lambda = - \frac{32}{3} \pi T_{11}
= - \frac{32}{3} \pi p\,,
\label{eqset4}
\ee

\be
- \frac{\partial}{\partial x} \left(\frac{\partial \beta}{\partial t}
+ \frac{1}{2} \frac{\partial}{\partial x} (\beta^2)\right) 
- 2 \Lambda = 8 \pi(T_{33} + T_{22}) = 16 \pi p\,,
\label{eqset6}
\ee

\be
\frac{\partial^2 \beta}{\partial x \partial y} = 
- 16 \pi (T_{02} + \beta T_{12}) = 0\,,
\label{eqset7}
\ee

\be
\frac{\partial^2 \beta}{\partial x \partial z} = 
- 16 \pi (T_{03} + \beta T_{13}) = 0\,,
\label{eqset8}
\ee

\be
\frac{\partial \beta}{\partial y}\frac{\partial \beta}{\partial z}
= 16 \pi T_{23} = 0\,\,,
\label{eqset9}
\ee

\subsection{Solving the Einstein equations with $\Lambda$ for the
perfect fluid}

From Eq. \eqref{eqset9} it follows that either $\partial \beta/ 
\partial z = 0$, or $\partial \beta/\partial y = 0$, or both vanish. 
From Eqs. \eqref{eqset7} and \eqref{eqset8} it is easy to see
that $\partial \beta/\partial x$ can also be zero. Those 
cases reveals four possibilities, which we will discuss in detail 
as follows.
\begin{description}[align=left]

\item[Case 1: \small $\bm{\left[\displaystyle\frac{\partial\beta}
	{\partial z}=0\right]}$]

\item[Case 1a: \small $\bm{\left[\displaystyle \frac{\partial
     \beta}{\partial z} = 0 \,\,\,\, \text{and} \,\,\,\, \frac{\partial
     \beta}{\partial x} = 0\right]}$]
For this case Eqs.\ \eqref{eqset1} to \eqref{eqset9} simplify to,
\be
\Lambda =  6 \pi \left(\mu - \frac{p}{3}\right) \,,	
\label{eqset1case1}
\ee

\be
\left(\frac{\partial\beta}{\partial y}\right)^2  
 = 4 (\Lambda - 8 \pi \mu)\,\,,
\label{eqset2case1a}
\ee

\be
\left(\frac{\partial \beta}{\partial y}\right)^2 
= \frac{4}{3}\pi(\Lambda - 8 \pi p)\,\,.
\label{eqset3case1a}
\ee

The set of these last equations implies that the shift vector 
$\beta$ is not uniquely defined.   It is a complex valued
function that depends only on $(t,y)$ spacetime coordinates.
In the case of the dust EMT as $p = 0$, the warp drive metric
is no longer a vacuum solution as it was found in \cite{nos},
because the existence of the cosmological constant as another
parameter originated a solution that does not consider shock waves
via the Burgers equation.

\item[Case 1b: \small $\bm{\left[\displaystyle \frac{\partial
     \beta}{\partial z} =0 \,\,\,\, \text{and} \,\,\,\, \frac{\partial
     \beta}{\partial y} =0\right]}$]
 
For this case one has to solve the following equations,
\be
\Lambda = 8 \pi \mu = 8 \pi p = 0 \,,	
\label{eqset1case1b}
\ee

\be
- \frac{\partial}{\partial x} \left(\frac{\partial \beta}{\partial t}
+ \frac{1}{2} \frac{\partial}{\partial x} (\beta^2)\right) = 0\, .
\label{eqset2case1b}
\ee

Eq.\ \eqref{eqset2case1b} is the Burgers equation that connects the 
warp drive to shock waves, as discussed in Ref.\ \cite{nos}. The
cosmological constant, fluid pressure and matter density are all equal
to zero and the warp drive metric \eqref{alcmetric1} is a vacuum
solution for the Einstein equations.

\item[Case 2: \small $\bm{\left[\displaystyle\frac{\partial\beta}
	{\partial y}=0\right]}$]
	
\item[Case 2a: $\bm{\displaystyle \left[\frac{\partial\beta}{\partial 
y} = 0 \ \text{and} \ \frac{\partial \beta}{\partial x} = 0\right]}$]
		
For this configuration, the set of Eqs. \eqref{eqset1} to 
\eqref{eqset9} simplify to
\be
\Lambda = 6 \pi \left(\mu - \frac{p}{3}\right) \,,	
\label{eqset1case2a}
\ee

\be
\left(\frac{\partial\beta}{\partial z}\right)^2  
 = 4 (\Lambda - 8 \pi \mu)\,\,,
\label{eqset2case2a}
\ee

\be
\left(\frac{\partial \beta}{\partial z}\right)^2 
= \frac{4}{3}\ (\Lambda - 8 \pi p)\,.
\label{eqset3case2a}
\ee

The above set of equations are very similar to Case 1a, where 
the shift vector is a complex valued function and it is not uniquely 
defined, but in this case $\beta$ depends on the $(t,z)$ coordinates.

\item[Case 2b: \small $\bm{\left[\displaystyle \frac{\partial
     \beta}{\partial z} =0 \,\,\,\, \text{and} \,\,\,\, \frac{\partial
     \beta}{\partial y} =0\right]}$]
 
For this case, one has to solve the following equations
\be
\Lambda = 8 \pi \mu = 8 \pi p = 0\,,	
\label{eqset1case2b}
\ee

\be
- \frac{\partial}{\partial x} \left(\frac{\partial \beta}{\partial t}
+ \frac{1}{2} \frac{\partial}{\partial x} (\beta^2)\right) = 0\,\,,
\label{eqset2case2b}
\ee
which is the same case as Case 1b.

\end{description}


\subsection{Divergence for the perfect fluid EMT} \label{divemts}

Calculating the divergence for the perfect fluid EMT, one arrives at 
the following equations,
\be
{T^{0 \nu}}_{;\nu} = - (\mu + p) \frac{\partial \beta}{\partial x}
- \beta \frac{\partial (p + \mu)}{\partial x} 
- \frac{\partial \mu}{\partial t} \,,
\label{nulldiv0}
\ee

\be
{T^{1 \nu}}_{;\nu} = \frac{\partial p}{\partial x} \,,
\label{nulldiv1}
\ee

\be
{T^{2 \nu}}_{;\nu} = \frac{\partial p}{\partial y} \,,
\label{nulldiv2}
\ee

\be
{T^{3 \nu}}_{;\nu} = \frac{\partial p}{\partial z} \,.
\label{nulldiv3}
\ee
Besides, imposing the null divergence condition, Eqs. \eqref{nulldiv0}
to \eqref{nulldiv3} implies that the pressure $p$ does not depend
on the spatial coordinates. Considering cases 1a and 2a there is
another partial differential equation to solve,
\be
\beta \frac{\partial \mu}{\partial x} 
+ \frac{\partial \mu}{\partial t} = 0\,,
\label{nulldiv01a2a}
\ee
and for Cases 1b and 2b Eq.\,\eqref{nulldiv0} is trivially
satisfied since $\mu = p = 0$.

\section{Energy conditions} \label{engconds}
\renewcommand{\theequation}{4.\arabic{equation}}
\setcounter{equation}{0}

\subsection{Weak Energy Conditions} 

For this case the EMT at each point of the spacetime must obey the
inequality
\be
T_{\alpha \sigma} \, u^\alpha u^\sigma \geq 0\,\,    
\label{weccond}
\ee
for any timelike vector $\textbf{u} \, (u_\alpha u^\alpha < 0)$ and 
any null zero vector $\textbf{k} \, (k_\alpha k^\alpha = 0)$. For an
observer with unit tangent vector $\textbf{v}$ at a certain point of
the spacetime, the local energy density measured by any observer is
non-negative \cite{HawkingEllis1973}. For the perfect fluid EMT the 
expression $T_{\alpha \sigma} \, u^\alpha u^\sigma$ is
\be
T_{\alpha \sigma} \, u^\alpha u^\sigma = \mu\,\,,    
\ee
and the weak energy condition from Eq.\,\eqref{weccond} is satisfied
if the matter density $\mu$ is positive. This is also the case for 
the dust EMT.

\subsection{Dominant Energy Conditions}

For every timelike vector $u_\alpha$, the following inequality must 
be satisfied,
\be
T^{\alpha \beta} \, u_\alpha u_\beta \geq 0, \quad \text{and} \quad 
F^\alpha  F_\alpha  
\leq 0\,\,, 
\ee
where $F^\alpha = T^{\alpha \beta} u_\beta$ is a non-spacelike 
vector, and the following condition must also be satisfied  
\be
T^{00} \geq |T^{\alpha \beta}|, \ \text{for each} \ \alpha
, \beta\,\,.
\ee

Evaluating the first condition for the perfect fluid EMT we have that,
\be
T^{\alpha \beta} \, u_\alpha u_\beta = \mu.
\label{dec1}
\ee
The other condition $F^\alpha  F_\alpha$ is given by the result
\be
F^\alpha  F_\alpha = - \mu^2 \leq 0\,\,.
\label{dec2}
\ee
Hence, the dominant energy condition is satisfied for $\mu > 0$, as
can be seen in Eq.\ \eqref{dec1}. Besides, Eq.\,\eqref{dec2} is 
always satisfied no matter the sign of the matter density. This
condition also holds true if one considers the dust EMT as a 
particular case for the perfect fluid with null pressure.

\subsection{Strong Energy Conditions}

For the strong energy condition the expression
\be
\Bigg(T_{\alpha \beta} - \frac{1}{2}T \, g_{\alpha \beta} \Bigg) 
u^\alpha u^\beta \geq 0 
\label{seccond}
\ee
is true for any timelike vector $u$. Computing the strong energy condition 
in Eq.\,\eqref{seccond} yields, 
\be
\Bigg(T_{\alpha \beta} - \frac{1}{2}T \, g_{\alpha \beta} \Bigg) 
u^\alpha u^\beta = \frac{1}{2}(3p + \mu)\,\,,
\ee
and the strong energy condition stated in Eq.\,\eqref{seccond} is
satisfied if $3p + \mu \geq 0$. The same is valid for the dust EMT,
considering $p=0$ for the perfect fluid, if $\mu \geq 0$.

\subsection{Null Energy Conditions}

The null energy conditions are satisfied in the limit of null 
observers. For the null vector $\textbf{k}$ the following conditions
must be satisfied,
\be
T_{\alpha \sigma} \, k^\alpha k^\sigma \geq 0, \qquad \text{for any null 
vector} \ k^\alpha\,\,.
\ee
Assuming that the following null vector $k^\alpha$ is given by,
\be
k^\alpha = (a,b,0,0)\,\,,    
\ee
we have that the relation between the components $a$ and $b$ are obtained by 
solving $k_\alpha k^\alpha = 0$. The two solutions given by,
\be
a = \frac{b}{\beta + 1} \qquad \mbox{and} \qquad  
a = \frac{b}{\beta - 1}\,\,.
\label{abnullvecsolved}
\ee
Then, the null energy condition reads,
\be
T_{\alpha \sigma} \, k^\alpha k^\sigma = 
\left(\frac{b}{\beta \pm 1}\right)^2
\left(\mu + p\right) \,,
\label{nec1}
\ee
and the null energy condition may be satisfied if the
following conditions are written as
\be
\mu + p \geq 0 \,\,.
\label{nec2}
\ee

Eq.\,\eqref{nec2} is also true for the dust EMT if one considers
it as a particular case for the perfect fluid with zero pressure,
then the null energy condition is satisfied for the dust if the
matter density is positive.

\section{Dust as a particular case from the perfect fluid}

Table \ref{tableenergycond} summarizes the results found for the
energy conditions for the perfect fluid with the cosmological constant
that are widely known \cite{HawkingEllis1973}. Considering the dust
EMT as a particular case for the perfect fluid by imposing the pressure
$p$ to be zero, the energy conditions for the warp drive metric and
the dust EMT would be trivially satisfied, since for this case, the
solution of the Einstein equations is a vacuum solution \cite{nos}.

\begin{table}[ht]
\caption{Summary results for the perfect fluid energy conditions} 
\centering 
\begin{tabular}[c]{l @{\hspace{50pt}} l} 
\hline\hline 
Energy condition & Results \\ [0.5ex] 
\hline 
Weak         &  $\mu \geq 0$  \\ 
Strong       &  $\mu \geq 0$  \\
Dominant     &  $\mu + 3p \geq 0$  \\
Null         &  $\mu + p \geq 0$ \\ 
[1ex] 
\hline 
\end{tabular}
\label{tableenergycond} 
\end{table}

Table \ref{tab1} summarizes the solutions of the Einstein equations for
the perfect fluid EMT with the cosmological constant and the warp drive
metric. As can be seen there are two types of solutions and each is
divided in two sub cases. Solutions 1b and 2b are identical and require
that $\Lambda=p=\mu=0$, where the two solutions are the ones already found
in Ref.\ \cite{nos} for the dust of non interacting particles EMT. This
led to a vacuum solution of the Einstein equations and the connection
between shock waves and the warp drive via the Burgers equation.

Solutions 1a and 2a in table \ref{tab1} have structures very similar to
the ones of the same type of equations, but for the solution 1a the shift
vector is a function of both the time and the $y$-spatial coordinates, i.e.,
$\beta = \beta(y,t)$. For solution 2a it is a function of both the time
and the $z$-spatial coordinate, i.e., $\beta = \beta(z,t)$. 

If we consider the dust solution as a particular case of  perfect 
fluid with the imposition that the pressure is zero, we have that the four 
sets of partial differential equations in Table \ref{tab1} become 
a solution for the warp drive metric and the dust EMT with the 
cosmological constant. In the case of dust EMT there is no longer a set 
of equations 1a and 2a to be solved, only 1b and 2b, that are identical
to the ones appearing in Ref.\ \cite{nos}. Even with a cosmological 
constant the dust EMT seems to be not a stable source of matter, energy
and momentum for the warp drive.

\begin{table}
\begin{tabular}{| m{3cm} | m{3cm} | m{8cm} |}
\hline 
Case & Condition & Results \\ 
\hline 
\multirow{2}{*}{$1) \
\displaystyle{\frac{\partial \beta}{\partial z} = 0}$}
& 
$1a) \ \displaystyle{\frac{\partial \beta}{\partial x} = 0}$
&
$\begin{array} {ll} 
\Lambda =  6 \pi \left(\mu - \frac{p}{3}\right) \\ [6pt]
\beta = \beta(y,t)\\ [6pt]
\displaystyle{\frac{\partial \beta}{\partial y} 
= \pm\sqrt{4 (\Lambda - 8 \pi \mu)}} \\ [8pt]
\displaystyle{\frac{\partial \beta}{\partial y} 
= \pm\sqrt{\frac{4}{3} (\Lambda - 8 \pi p)}} \\ [8pt]
\displaystyle{\beta \frac{\partial \mu}{\partial x} 
+ \frac{\partial \mu}{\partial t} = 0 
\ \ \text{(null divergence)}} \\ [8pt]
\end{array}$ \\ [28pt]
\cline{2-3}   
&
$1b) \ \displaystyle{\frac{\partial \beta}{\partial y} = 0}$
&
$\begin{array} {ll} 
\Lambda = 8 \pi \mu = 8 \pi p = 0\\ [6pt]
\beta = \beta(x,t)\\ [6pt]
\displaystyle{
\frac{\partial \beta}{\partial t}
+ \frac{1}{2} \frac{\partial}{\partial x} 
(\beta^2) = h(t)\, } \\ [6pt]
\text{Null divergence is trivially satisfied}\\ [2pt]
\text{This is the solution found in Ref.\ \cite{nos}} \\ [8pt]
\end{array}$ \\ [28pt] 
\hline 

\multirow{2}{*}{$2) \
\displaystyle{\frac{\partial \beta}{\partial y} = 0}$}
& 
$2a) \ \displaystyle{\frac{\partial \beta}{\partial x} = 0}$
&
$\begin{array} {ll} 
\Lambda =  6 \pi \left(\mu - \frac{p}{3}\right) \\ [6pt]
\beta = \beta(y,t)\\ [6pt]
\displaystyle{\frac{\partial \beta}{\partial z} 
= \pm\sqrt{4 (\Lambda - 8 \pi \mu)}} \\ [8pt]
\displaystyle{\frac{\partial \beta}{\partial z} 
= \pm\sqrt{\frac{4}{3} (\Lambda - 8 \pi p)}} \\ [8pt]
\displaystyle{\beta \frac{\partial \mu}{\partial x} 
+ \frac{\partial \mu}{\partial t} = 0 
\ \ \text{(null divergence)}} \\ [8pt]
\end{array}$ \\ [28pt]
\cline{2-3}   
&
$2b) \ \displaystyle{\frac{\partial \beta}{\partial z} = 0}$
&
$\begin{array} {ll} 
\Lambda = 8 \pi \mu = 8 \pi p = 0 \\ [6pt]
\beta = \beta(x,t)\\ [6pt]
\displaystyle{\frac{\partial \beta}{\partial t} 
+ \frac{1}{2} \frac{\partial}{\partial x}(\beta^2)
= h(t)} \\ [6pt]
\text{Null divergence is trivially satisfied}\\ [2pt]
\text{This is the solution found in Ref.\ \cite{nos}} \\ [8pt]
\end{array}$ \\ [28pt] 
\hline 
\end{tabular}
\caption{Summary of all solutions of the Einstein equation with
the cosmological constant and the Alcubierre warp drive metric 
having the perfect fluid EMT as mass-energy source. This table
is also valid for the dust particle if considered as a particular
case for the perfect fluid with null pressure.}
\label{tab1}
\end{table}

\section{Conclusions and final remarks}
\renewcommand{\theequation}{5.\arabic{equation}}
\setcounter{equation}{0}

In this work we investigated how the presence of a cosmological constant
would affect the solutions of the Einstein equations endowed with the
Alcubierre warp drive metric and the perfect fluid EMT as the source. 
Firstly, we solved the Einstein equations and obtained two solutions,
Cases 1b and 2b, that are similar to the solutions we found for the dust 
particle without the cosmological constant \cite{nos}, and two other
solutions, Case 1a with $\beta=\beta(t,y)$ and Case 2a with
$\beta=\beta(t,z)$, having the following equation of state relating the
cosmological constant $\Lambda$, the matter density $\mu$ and the fluid
pressure $p$: $\Lambda = 2 \pi (3 \mu - p).$

The presence of the cosmological constant allows the shift vector to be
a real valued function as can be seen from Eqs.\ \eqref{eqset2case1a} 
and \eqref{eqset3case1a} for Case 1a in Table \ref{tab1}, and Eqs.\
\eqref{eqset2case2a} and \eqref{eqset3case2a} for Case 2a, namely,
$\Lambda - 8\pi \mu \geq 0$ and $\Lambda - 8\pi p \geq 0.$

If we do not consider the cosmological constant, then the shift vector
would become a complex valued function for Cases 1a and 2a in Table
\ref{tab1}. The energy conditions are all satisfied for the perfect fluid
if the conditions in Table \ref{tableenergycond} are satisfied. Solutions
1b and 2b shown in Table \ref{tab1} connect the Burgers equation to both
the warp drive and the perfect fluid solution. In Ref.\ \cite{nos} we
found this intrinsic relationship between the warp drive and shock waves
by solving Einstein equations for the warp drive metric and the dust 
particle EMT, but we concluded that there is an impossibility of coupling
the dust as a source in this case. So, the presence of shock waves would
imply that the Alcubierre metric shown in Eq.\ \eqref{alcmetric1} is a
vacuum solution for the warp drive. In Ref.\ \cite{Alcubierre1994} the
Einstein equations were not solved, since the metric was merely guessed
with a form function (see Eq.\ \ref{regfunction}) that rules the warp
bubble shape. 

The results found here led us to a kind of prescription where the
warp drive requires more complex forms of matter than dust in order to
obtain stable solutions. In addition, considering this work and the
previous ones of this series of papers \cite{nos,nos2,nos3} it becomes
increasingly clear and that negative matter density may not be a
strict requirement to obtain warp speeds. The shift vector in the
direction of the warp bubble movement creates a coupling in the Einstein
equations that requires off-diagonal terms in the EMT source. In the
light of these results we may conjecture that the key for engineering a
superluminal propelling system for interstellar travel could be
understood as a complex distribution of energy, matter and momentum
sources that could stabilize the warp drive geometry, allowing then
superluminal travel.

\section*{Acknowledgments}

\ni E.M.C.A.\ thanks CNPq (Conselho Nacional de Desenvolvimento 
Cient\'ifico e Tecnol\'ogico), Brazil's federal scientific supporting 
agency, for partial financial support: grant number 406894/2018-3. 


\end{document}